\documentclass[showkeys,showpacs,superscriptaddress]{revtex4-2}
\usepackage{amsmath}
\usepackage{amssymb}
\usepackage{graphicx}
\usepackage{float}
\usepackage{xcolor}
\usepackage{physics}
\usepackage{ulem}

\allowdisplaybreaks

\begin{document}

\title{
Axially symmetric particlelike solutions with the flux of a magnetic field in the non-Abelian Proca-Higgs theory
}

\author{
Vladimir Dzhunushaliev
}
\email{v.dzhunushaliev@gmail.com}
\affiliation{
Department of Theoretical and Nuclear Physics,  Al-Farabi Kazakh National University, Almaty 050040, Kazakhstan
}
\affiliation{
Institute of Nuclear Physics, Almaty 050032, Kazakhstan
}
\affiliation{
Academician J.~Jeenbaev Institute of Physics of the NAS of the Kyrgyz Republic, 265 a, Chui Street, Bishkek 720071, Kyrgyzstan
}

\author{Vladimir Folomeev}
\email{vfolomeev@mail.ru}
\affiliation{
Institute of Nuclear Physics, Almaty 050032, Kazakhstan
}
\affiliation{
Academician J.~Jeenbaev Institute of Physics of the NAS of the Kyrgyz Republic, 265 a, Chui Street, Bishkek 720071, Kyrgyzstan
}
\affiliation{
International Laboratory for Theoretical Cosmology, Tomsk State University of Control Systems and Radioelectronics (TUSUR),
Tomsk 634050, Russia
}

%\date{\today}

\begin{abstract}
Within the non-Abelian SU(2) Proca-Higgs theory, we study localised axially symmetric solutions possessing a finite field energy.
It is shown that in a certain sense such solutions are analogues of the Nielsen-Olesen tube,
since they have a longitudinal magnetic field creating a flux of this field over the central cross-section of the Proca tube.
The main difference between the Proca tube and the Nielsen-Olesen tube is that the Proca tube is described by a topologically trivial solution and has
finite size, since its energy density decreases exponentially with distance.
The dependence of the total field mass of the Proca tube on the value of one of the parameters determining the solution is examined in detail.
The solutions are obtained both in the presence and in the absence of external sources (charge and current densities).
\end{abstract}

\pacs{11.90.+t
}

\keywords{non-Abelian Proca theory, Higgs field, finite-size flux tube solutions
}
\date{\today}

\maketitle

\section{Introduction}

In recent years interest in systems involving various massive vector fields has increased considerably.
Such fields do emerge, in particular, when considering Proca theories. The presence of a mass of vector fields results in
substantial differences in the behaviour of Proca fields compared with Yang-Mills fields: the Proca fields decay exponentially
fast  at spatial infinity, which is impossible in principle for the Yang-Mills fields.

At the present time, studies of Proca theories are performed in various directions. In particular,
within the past few years, there was prolific activity in the field of physics of
Proca stars~\cite{Brito:2015pxa,Herdeiro:2017fhv,Dzhunushaliev:2019kiy,Herdeiro:2019mbz,Bustillo:2020syj}
and black holes modelled within generalized Proca theories~\cite{Heisenberg:2017xda}.
The massive extension of a SU(2) gauge theory is studied in Refs.~\cite{Allys:2015sht,Allys:2016kbq},
the static and spherically symmetric solutions in a class of the generalized
Proca theory with the nonminimal coupling to the Einstein tensor are found in Ref.~\cite{Minamitsuji:2016ydr},
the cosmological implications of generalized Proca theories are under investigation in Ref.~\cite{DeFelice:2016yws}.

In Refs.~\cite{Dzhunushaliev:2019sxk,Dzhunushaliev:2020eqa,Dzhunushaliev:2021uit}, we have studied infinite cylindrically symmetric solutions (tubes)
containing a longitudinal electric field. The solutions have been obtained within the non-Abelian Proca-Higgs theory.
It is of great interest to get similar solutions describing finite-size objects; this implies that the field energy density should decrease away from the center of
a configuration sufficiently fast. In this connection,
in the present study we extend the results of Refs.~\cite{Dzhunushaliev:2019sxk,Dzhunushaliev:2020eqa,Dzhunushaliev:2021uit} and
examine axially symmetric solutions in the non-Abelian Proca-Higgs theory with a longitudinal color magnetic field
directed along the symmetry axis and transverse electric fields located in a plane perpendicular to the symmetry axis.
As will be shown below, the asymptotic behavior of the fields leads to the fact that the total energy of such configurations
is finite; this enables us to call such solutions as particlelike solutions.
At the same time, the field configurations obtained have a non-Abelian longitudinal magnetic Proca field; this enables us to call them tubes with a flux of the
corresponding magnetic field over a plane perpendicular to the tube axis and located at the center of the axis. The latter means that such a tube can be regarded as an analogue of
the Nielsen-Olesen tube~\cite{NielsenOlesen:1973}, since in both cases there is the flux of the magnetic field over the transverse cross-section of the tube.
On the other hand, there also exist substantial differences:
(i)~the Nielsen-Olesen tube is a topologically nontrivial configuration, whereas our field configuration is topologically trivial;
(ii)~the Proca tube has finite size, since the fields decrease exponentially with distance; and
(iii)~the Nielsen-Olesen tube contains an Abelian magnetic field, whereas the Proca tube~-- a non-Abelian magnetic field.

It should be also noted that in our study we consider a system supported by Proca fields together with a Higgs scalar field. The presence of the latter, as will be shown below,
leads to a very interesting result:  it initiates the process where the Proca fields are being pushed outside by the Higgs field.
This interesting result is an analogue of the Meissner phenomenon in superconductivity and of the dual Meissner effect in QCD.

The field configurations under investigation can emerge as quasiparticles in the hypothetical ``quark-Proca-gluon-Higgs'' plasma.
In this connection we note that, in QCD, it is assumed that in the quark-gluon plasma there exist multiple bound states of quasiparticles
(for details see Ref.~\cite{Shuryak:2003ty} where the existence of such quasiparticles has been suggested, as well as Ref.~\cite{Shuryak:2004tx}
where studies in this direction have been continued and the role of multiple (colored) bound states in the quark-gluon plasma phase at not too high temperature is demonstrated).

Thus, we will study here particlelike solutions in the Proca-Higgs theory in which there is a color longitudinal Proca field directed along the symmetry axis;
this enables us to call such solutions as Proca tubes with a flux of the magnetic field.
The paper is organized as follows. In Sec.~\ref{Proca_Dirac_scalar}, we write down the general field equations for the non-Abelian-Proca-Higgs theory.
In Sec.~\ref{inf_tube}, we obtain cylindrically symmetric solutions to the equations of Sec.~\ref{Proca_Dirac_scalar} describing infinite tubes with
the flux  of the longitudinal color magnetic field. In Sec.~\ref{fin_tube}, we find axially symmetric solutions to the equations of Sec.~\ref{Proca_Dirac_scalar}
(both with and without charge and current densities) describing finite-size tubes with the flux of the longitudinal chromomagnetic field.
Finally, in Sec.~\ref{concl}, we summarize and discuss the results obtained in the present paper.

\section{Non-Abelian SU(3) Proca-Higgs theory}
\label{Proca_Dirac_scalar}

The Lagrangian describing a system consisting of a non-Abelian SU(3) Proca field $A^a_\mu$ interacting with nonlinear scalar field $\phi$ can be taken in the form
(hereafter, we work in units such that $c=\hbar=1$)
\begin{equation}
	\mathcal L =  - \frac{1}{4} F^a_{\mu \nu} F^{a \mu \nu} -
	\frac{\left( \mu^2 \right)^{a b, \mu}_{\phantom{a b,}\nu}}{2}
	A^a_\mu A^{b \nu} +
	\frac{1}{2} \partial_\mu \phi \partial^\mu \phi +
	\frac{\lambda}{2} \phi^2 A^a_\mu A^{a \mu} -
	\frac{\Lambda}{4} \left( \phi^2 - M^2 \right)^2.
\label{0_10}
\end{equation}
Here
$
	F^a_{\mu \nu} = \partial_\mu A^a_\nu - \partial_\nu A^a_\mu +
	g f_{a b c} A^b_\mu A^c_\nu
$ is the field strength tensor for the Proca field, where $f_{a b c}$ are the SU(3) structure constants, $g$ is the coupling constant,
$a,b,c = 1,2, \dots, 8$ are color indices,
$\mu, \nu = 0, 1, 2, 3$ are spacetime indices. The Lagrangian \eqref{0_10} also contains the arbitrary constants $M, \lambda, \Lambda$ and the Proca field mass matrix
$
	\left( \mu^2 \right)^{a b, \mu}_{\phantom{a b,}\nu}
$.

Using \eqref{0_10}, the corresponding field equations can be written in the form
\begin{align}
	D_\nu F^{a \mu \nu} - \lambda \phi^2 A^{a \mu} =&
	- \left( \mu^2 \right)^{a b, \mu}_{\phantom{a b,}\nu} A^{b \nu}+j^{a\mu},
\label{0_20}\\
	\Box \phi =& \lambda A^a_\mu A^{a \mu} \phi +
	\Lambda \phi \left( M^2 - \phi^2 \right) ,
\label{0_30}
\end{align}
where, for the sake of generality, we have also added the current four-vector $j^{a\mu}$.
For such systems, the field energy density is
\begin{equation}
\begin{split}
	\varepsilon = &\frac{1}{2} \left( E^a_i \right)^2 +
	\frac{1}{2} \left( H^a_i \right)^2 -
	\left[
		\left( \mu^2 \right)^{a b, \alpha}_{\phantom{a b,} 0} A^a_\alpha A^b_0 -
		\frac{1}{2} \left( \mu^2 \right)^{a b, \alpha}_{\phantom{a b,} \beta} A^a_\alpha A^{b \beta}
	\right]
	+
	\frac{1}{2} \left( \partial_t \phi \right)^2 +
	\frac{1}{2} \left( \nabla \phi \right)^2
\\
	&
	+\lambda \phi^2 \left[
		\left( A^a_0 \right)^2 - \frac{1}{2} A^a_\alpha A^{a \alpha}
	\right] +
	\frac{\Lambda}{4} \left( \phi^2 - M^2 \right)^2 ,
\label{0_40}
\end{split}
\end{equation}
where $i=1,2,3$ and $E^a_i$ and $H^a_i$ are the components of the electric and magnetic field strengths, respectively.

In the present paper we will consider solutions belonging to the subgroup  $\text{SU(2)}\subset \text{SU(3)}$
spanned on the Gell-Mann matrices $\lambda^{2,5,7}$.

\section{Infinite flux tube solutions}
\label{inf_tube}

To begin with, it will be helpful to study a simpler system with an infinite flux tube containing a
flux of a longitudinal electric field~\cite{Dzhunushaliev:2019sxk,Dzhunushaliev:2020eqa,Dzhunushaliev:2021uit}.
To describe such a tube, let us choose the ansatz for the field potentials in the form
\begin{equation}
	A^2_t = \frac{f(\rho)}{g} , \;
	A^7_\varphi = \frac{\rho w(\rho)}{g} , \;
 	\phi = \phi(\rho)
\label{1_10}
\end{equation}
written in cylindrical coordinates $\{t,\rho, z, \varphi\}$. Such a tube contains the following color electric and magnetic fields (physical components):
\begin{equation}
E^2_\rho =  -\frac{f^\prime}{g} , \quad E^5_\varphi = -  \frac{ f w}{2 g} , \quad
H^7_z =  - \frac{\rho w^\prime + w }{g \rho }.
\label{elec_magn_str}
\end{equation}
(Henceforth in this section the prime denotes differentiation with respect to  $\rho$.)
In this case the energy flux is absent, since all components of the Poynting vector are zero,
\begin{equation}
	S^i = \frac{\epsilon^{i j k}}{\sqrt{\gamma}} E^a_j H^a_k = 0,
\label{3_a_40}
\end{equation}
where $\epsilon^{i j k}$ is the completely antisymmetric Levi-Civita symbol and $\gamma$ is the determinant of the space metric.

Substituting the potentials \eqref{1_10} in Eqs.~\eqref{0_20} and \eqref{0_30}, we get the following set of equations (without the current):
\begin{eqnarray}
  f'' + \frac{ f'}{\rho} &=& f
  \left(
  	 \frac{ w^2}{4} + \lambda \phi^2 -
  	 \mu_1^2
  \right) ,
\label{3_a_93}\\
  w'' + \frac{w'}{x} - \frac{w}{\rho^2} &=&
   w \left(
  	- \frac{ f^2}{4}  +  \lambda \phi^2 -
  	 \mu_2^2
  \right),
\label{3_a_97}\\
  \phi'' + \frac{\phi'}{\rho} &=&
   \phi
  \left[
  	 \frac{\lambda}{g^2} \left( -  f^2 +   w^2 \right) +
  	 \Lambda \left( \phi^2 -  M^2 \right)
  \right]
\label{3_a_99}
\end{eqnarray}
with the following components of the Proca field mass matrix: $\mu_{1}^2=\left( \mu^2 \right)^{2 2, t}_{\phantom{a b,}t}$
 and $\mu_{2}^2=\left( \mu^2 \right)^{7 7, \varphi}_{\phantom{a b,}\varphi}$.
We seek a solution to Eqs.~\eqref{3_a_93}-\eqref{3_a_99} in the vicinity of the origin of coordinates in the form
\begin{eqnarray}
	 f(\rho) &=& f_0 +  f_2 \frac{\rho^2}{2} + \dots \quad \text{with} \quad
 f_2 = \frac{f_0}{2} \left[
		 \lambda \phi_0^2 -   \mu_1^2
	\right] ,
\label{3_a_100}\nonumber\\	
 w(\rho) &=&  w_1 \rho + \dots ,
\label{3_a_115}\nonumber\\
	\phi(\rho) &=&  \phi_0 +  \phi_2 \frac{\rho^2}{2} + \dots \quad \text{with} \quad
 \phi_2 = \frac{ \phi_0}{2 g^2} \left[
		- \lambda f_0^2 +
		g^2\Lambda \left(
			 \phi_0^2 -  M^2
	\right)
	\right],
\label{3_a_120}
\nonumber
\end{eqnarray}
where the expansion coefficients $f_0,  \phi_0$, and $w_1$ are arbitrary.

\begin{figure}[t]
\begin{minipage}[t]{.49\linewidth}
	\begin{center}
		\includegraphics[width=1\linewidth]{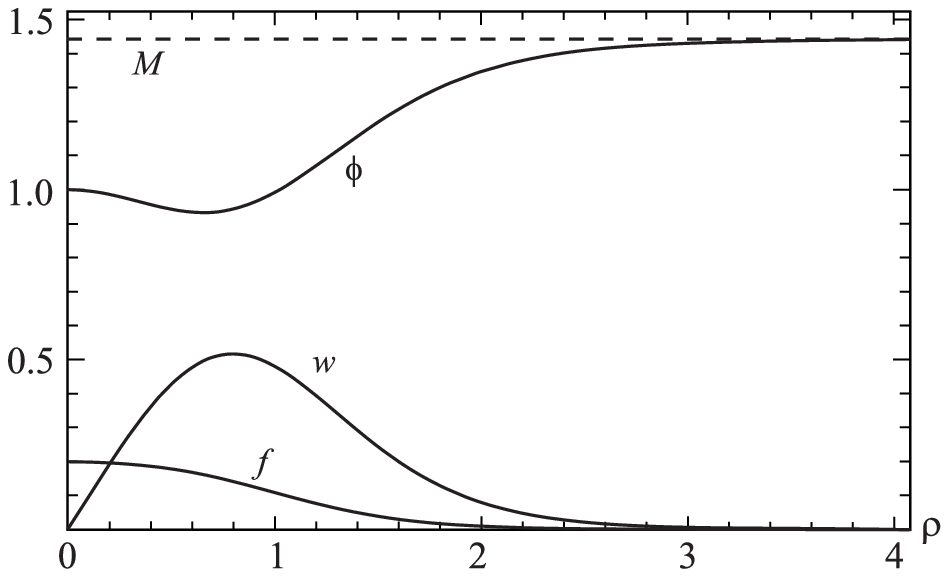}
	\end{center}
\vspace{-0.5cm}		
\caption{The graphs of the Proca field potentials
$f $, $w$ and of the scalar field $\phi$ for the following values of the system parameters:
$
	\lambda =10,  \Lambda = 1, g=1,  f_0 = 0.2,
	 \phi_0 = 1,  w_1 = 1,
	 M = 1.44227, \mu_1=3.359585, \mu_2=3.7633.
$
}
\label{f_w_phi}
\end{minipage}\hfill
\begin{minipage}[t]{.49\linewidth}
	\begin{center}
		\includegraphics[width=1\linewidth]{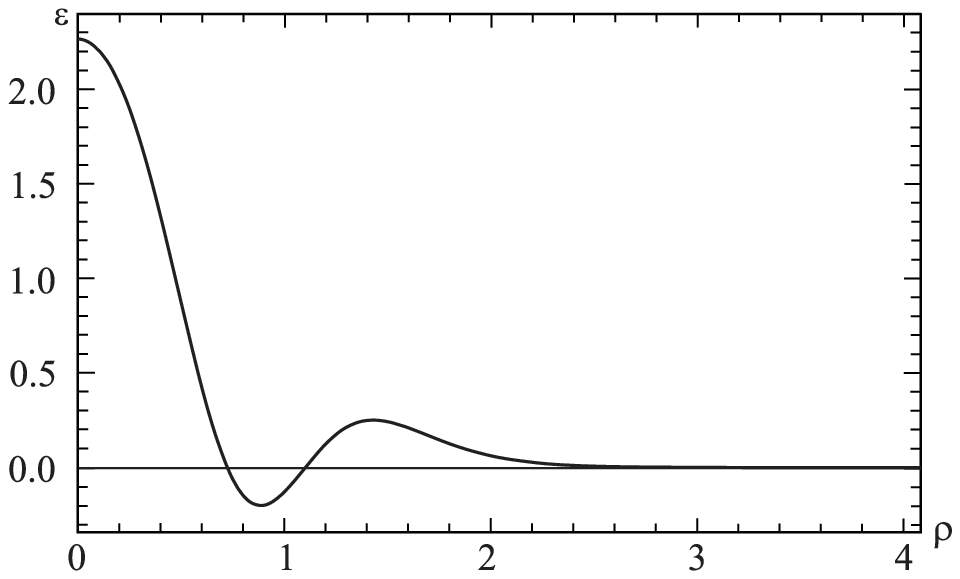}
	\end{center}
\vspace{-0.5cm}
\caption{The profile of the flux tube energy density from
 Eq.~\eqref{inf_energ_dens} for the solutions given in Fig.~\ref{f_w_phi}. 	
}
\label{energy_density}
\end{minipage} \\
\begin{minipage}[t]{.49\linewidth}
	\begin{center}
		\includegraphics[width=1\linewidth]{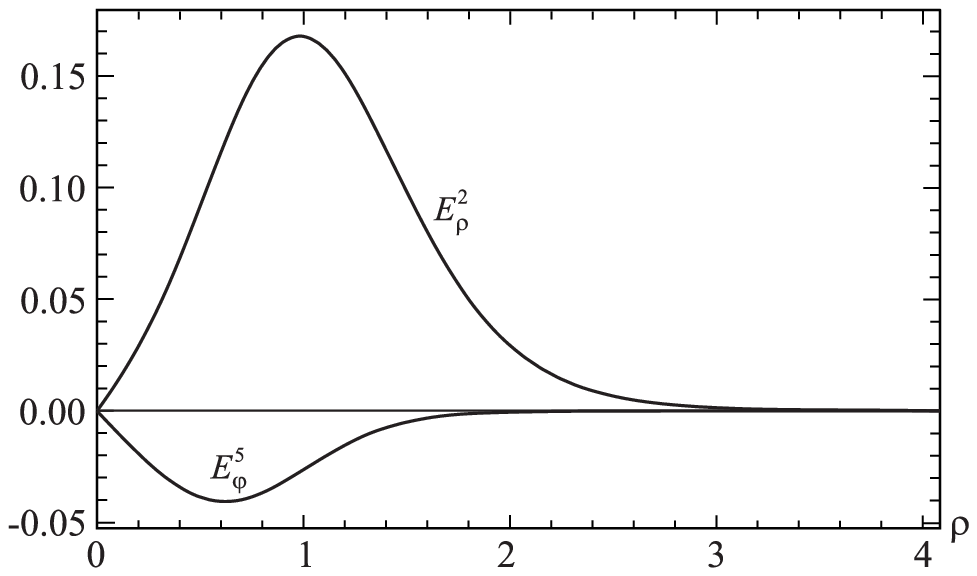}
	\end{center}
\vspace{-0.5cm}
\caption{The profiles of the color electric fields	
$ E^2_\rho$ and $E^5_\varphi $ from Eq.~\eqref{elec_magn_str} for the solutions given in Fig.~\ref{f_w_phi}.
}
\label{electric_fields}
\end{minipage} \hfill
\begin{minipage}[t]{.49\linewidth}
	\begin{center}
		\includegraphics[width=1\linewidth]{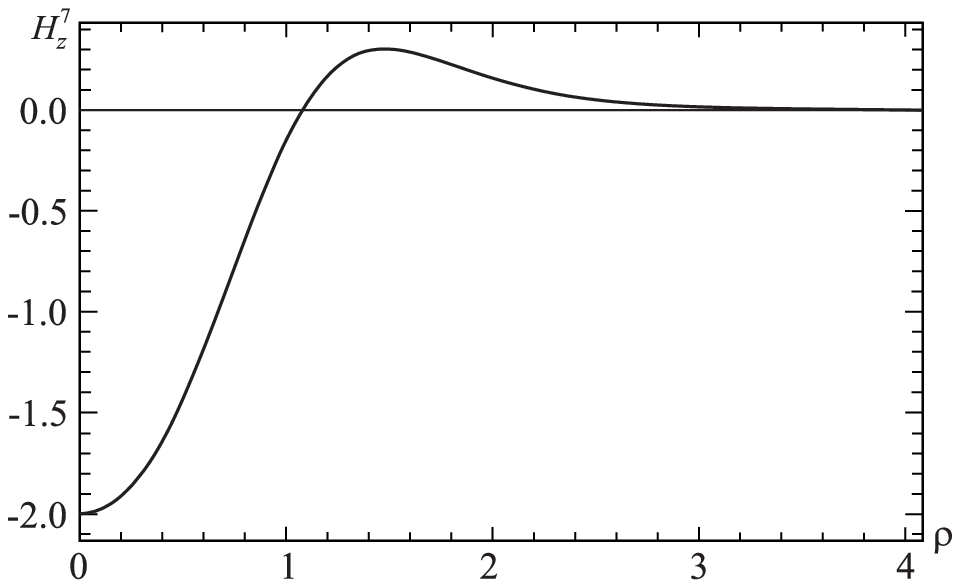}
	\end{center}
\vspace{-0.5cm}
\caption{The profile of the color magnetic field	
 $ H^7_z$ from Eq.~\eqref{elec_magn_str} for the solutions given in Fig.~\ref{f_w_phi}.
}
\label{magnetic_fields}
\end{minipage} \hfill
\end{figure}

The derivation of solutions to the set of equations~\eqref{3_a_93}-\eqref{3_a_99} is an eigenvalue problem
for the parameters $\mu_1, \mu_2$, and $ M$.
The numerical solution describing the behavior of the Proca field potentials and of the corresponding electric and magnetic fields is given in Figs.~\ref{f_w_phi}-\ref{magnetic_fields}.
In particular, Fig.~\ref{energy_density} shows the energy density obtained from Eq.~\eqref{0_40} using~\eqref{1_10} and~\eqref{elec_magn_str} in the form
\begin{equation}
\varepsilon=\frac{1}{g^2}\left[
\frac{f^{\prime 2}}{2}+\frac{w w^\prime}{\rho}+\frac{w^{\prime 2}}{2}+\frac{g^2}{2}\phi^{\prime 2}+
\frac{w^2}{2\rho^2}+\frac{f^2 w^2}{8}-\frac{\mu_1^2 f^2}{2}-\frac{\mu_2^2 w^2}{2}+
\frac{\lambda}{2}\left(f^2+w^2\right)\phi^2+\frac{g^2\Lambda}{4}\left(\phi^2-M^2\right)^2
\right].
\label{inf_energ_dens}
\end{equation}

The asymptotic behavior of the functions $f,  w$, and $\phi$, which follows from Eqs.~\eqref{3_a_93}-\eqref{3_a_99}, is
\begin{equation}
	 f(\rho) \approx  f_{\infty}
	\frac{e^{- \rho \sqrt{\lambda M^2 -  \mu_1^2}}}{\sqrt \rho}, \quad
	w(\rho) \approx  w_{\infty}
	\frac{e^{- \rho \sqrt{\lambda M^2 -  \mu_2^2}}}{\sqrt \rho},\quad
	 \phi \approx  M -  \phi_\infty
	\frac{e^{- \rho \sqrt{2 \Lambda  M^2}}}{\sqrt \rho} ,
\label{3_a_160}
\nonumber
\end{equation}
where $f_{\infty},   w_{\infty}$, and $\phi_\infty$ are integration constants.

It is seen from Figs.~\ref{f_w_phi}-\ref{magnetic_fields} that the
cylindrically symmetric solutions found can be used to describe infinitely long tubes within the
non-Abelian Proca-Higgs theory under consideration. Such tubes possess the finite linear energy density (see Fig.~\ref{energy_density}),
$$
	\mathcal{E} = 2 \pi \int \limits_0^\infty \rho\, \varepsilon(\rho) d \rho
	< \infty ,
$$
and the finite flux of the longitudinal color magnetic field $H^7_z$ (see Fig.~\ref{magnetic_fields}),
$$
	\Phi_z^H = 2 \pi \int \limits_0^\infty \rho H^7_z d \rho < \infty .
$$

Note that, unlike the systems considered by us earlier in Refs.~\cite{Dzhunushaliev:2019sxk,Dzhunushaliev:2020eqa,Dzhunushaliev:2021uit},
these tubes contain neither the flux of the longitudinal color
electric field $E^a_z$ nor the energy flux/momentum density (since the Poynting vector is zero).

\section{Finite flux tube solutions}
\label{fin_tube}

Let us now extend the solutions of the previous section to the case of finite-size configurations.
To obtain static axially symmetric solutions, we employ the ansatz~\eqref{1_10}, with an obvious generalization to the case where the components
of the four-potential depend also on the coordinate $z$. As will be demonstrated below, in this case, it is possible to get solutions describing configurations
possessing finite sizes both along the $\rho$-axis and along the $z$-axis.

For convenience of performing numerical computations, we employ spherical coordinates $\{t, r, \theta, \varphi\}$
in which the ansatz \eqref{1_10} can be recast in the form
\begin{equation}
	A^2_t = \frac{f(r, \theta)}{g} , \;
		A^7_\varphi = r\sin{\theta}\frac{w(r, \theta)}{g} , \;
 	\phi = \phi(r, \theta),
\label{vec_pot}
\end{equation}
where all functions now depend on
 $r$ and $\theta$. For such ansatz, there are the following nonzero physical components of color electric and magnetic fields:
\begin{align}
	E^2_r = & -\frac{f_{,r}}{g} , \quad
	E^2_\theta = -\frac{f_{, \theta}}{g r} , \quad
	E^5_\varphi = -  \frac{ f w}{2 g} ,
\label{2_10}\\
	H^7_r = &  -\frac{w_{, \theta} + w \cot \theta}{g r } , \quad
	H^7_\theta = \frac{w + r w_{,r}}{g r} .
\label{2_20}
\end{align}
(Henceforth a comma in lower indices denotes differentiation with respect to the corresponding coordinate.)
In turn, the field equations \eqref{0_20} and \eqref{0_30} yield
\begin{align}
	f_{, rr} + \frac{2 }{r}f_{, r} +
	\frac{1}{r^2} \left(
		f_{, \theta \theta} + \cot \theta f_{, \theta}
	\right) -
	\frac{1}{4} f \left(
		w^2 + 4	\lambda  \phi^2
	\right) +
	\mu_1^2 f = & j^{2 t}  ,
\label{2_30}\\
  w_{, rr} + \frac{2 }{r}w_{, r} + \frac{1}{r^2}
  \left(
  	w_{, \theta \theta} + \cot \theta w_{, \theta}
  \right) - \frac{w}{r^2\sin^2 \theta}+
	\frac{1}{4}\left(
	f^2 - 4 \lambda  \phi^2
	\right) w +
	\mu_2^2  w= &-r \sin{\theta} j^{7 \varphi}  ,
\label{2_40}\\
	 \phi_{, rr} + \frac{2}{ r }\phi_{, r}
	+ \frac{1}{r^2} \left(
		\phi_{, \theta \theta} + \cot \theta \phi_{, \theta}
	\right) +
	\frac{\lambda}{g^2}  \left(
		 f^2 - w ^2
	\right) \phi -
	 \Lambda \phi \left(\phi^2 - M^2\right) = & 0 .
\label{2_50}
\end{align}
The results of numerical simulations for these equations are given in Sec.~\ref{no_charge} (in the absence of the currents) and in Sec.~\ref{with_charge}
(in the presence of the currents).

\subsection{The case with no charges
}
\label{no_charge}

Consider first the case where the currents $j^{a \mu}=0$ in Eqs.~\eqref{2_30} and \eqref{2_40}.
To solve the set of equations \eqref{2_30}-\eqref{2_50}, it is necessary to impose
appropriate boundary conditions for the Proca and scalar fields at the origin ($r=0$), at infinity ($r\to \infty$), on the positive $z$ axis ($\theta=0$), and,
using the reflection symmetry with respect to $\theta\to \pi-\theta$, in the $\{x, y\}$ plane ($\theta=\pi/2$). So we require
\begin{align}
&\left. \frac{\partial f}{\partial r}\right|_{r = 0} =
	\left. \frac{\partial \phi}{\partial r}\right|_{r = 0} =  0,  \left. w \right|_{r = 0} = 0;
\quad\left. f \right|_{r = \infty} = 0 ,
	\left. w \right|_{r = \infty} = 0 ,
	\left. \phi \right|_{r = \infty} =  M ; \nonumber\\
&\left. \frac{\partial f}{\partial \theta}\right|_{\theta = 0} =
	\left. \frac{\partial \phi}{\partial \theta}\right|_{\theta = 0} =  0 ,  \left. w \right|_{\theta = 0} = 0 ;
\quad\left. \frac{\partial f}{\partial \theta}\right|_{\theta = \pi/2} =
	\left. \frac{\partial w}{\partial \theta}\right|_{\theta = \pi/2} =
	\left. \frac{\partial \phi}{\partial \theta}\right|_{\theta = \pi/2} =  0.\nonumber
\end{align}

\begin{figure}[h]
\includegraphics[width=.9\linewidth]{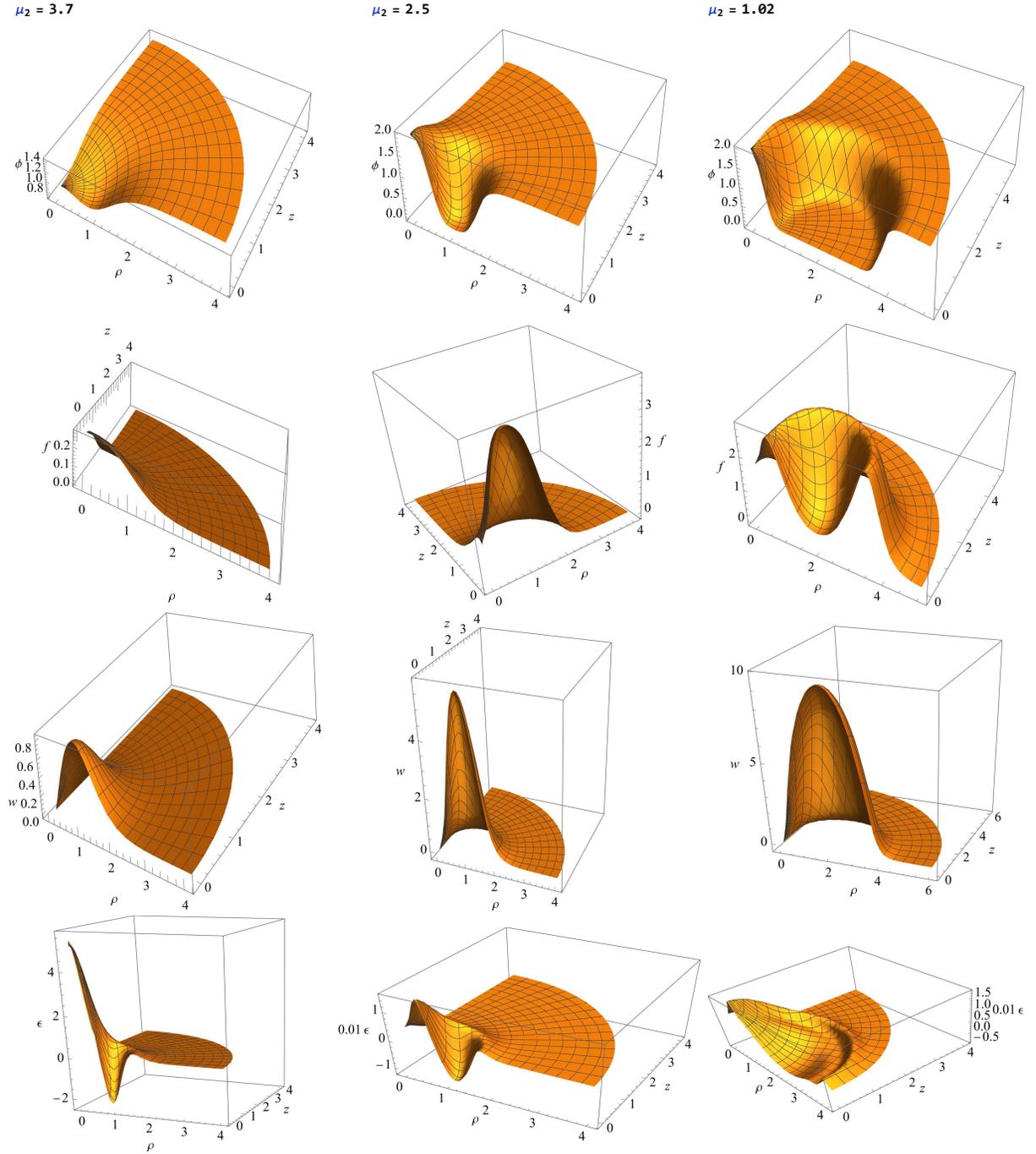}
%\vspace{-0.5cm}
\caption{Distributions of the fields $\phi$ (upper row), $f$ (second row),
$w$ (third row), and the total energy density of the system~$\varepsilon$ from Eq.~\eqref{2_80} (lower row)	for different
values of $\mu_2$ and fixed values of the parameters $M=1.4$, $\Lambda=1$, $\lambda=10$, $g=1$, and $\mu_1=3.3$.
}
\label{fig_fields_distr}
\end{figure}

The set of three coupled nonlinear elliptic partial differential equations \eqref{2_30}-\eqref{2_50}  has been
solved numerically subject to the above boundary conditions. Calculations are carried out using the package FIDISOL~\cite{fidisol}
with typical errors less than $10^{-4}$. As in the case of an infinite tube of Sec.~\ref{inf_tube}, the input parameters are eigenparameters
$\mu_1, \mu_2$, and $ M$, whose magnitudes determine  the solution completely. For different values of the aforementioned parameters,
one can obtain qualitatively different field distributions, some typical examples of which are depicted in Fig.~\ref{fig_fields_distr}.

\begin{figure}[h!]
\includegraphics[width=.5\linewidth]{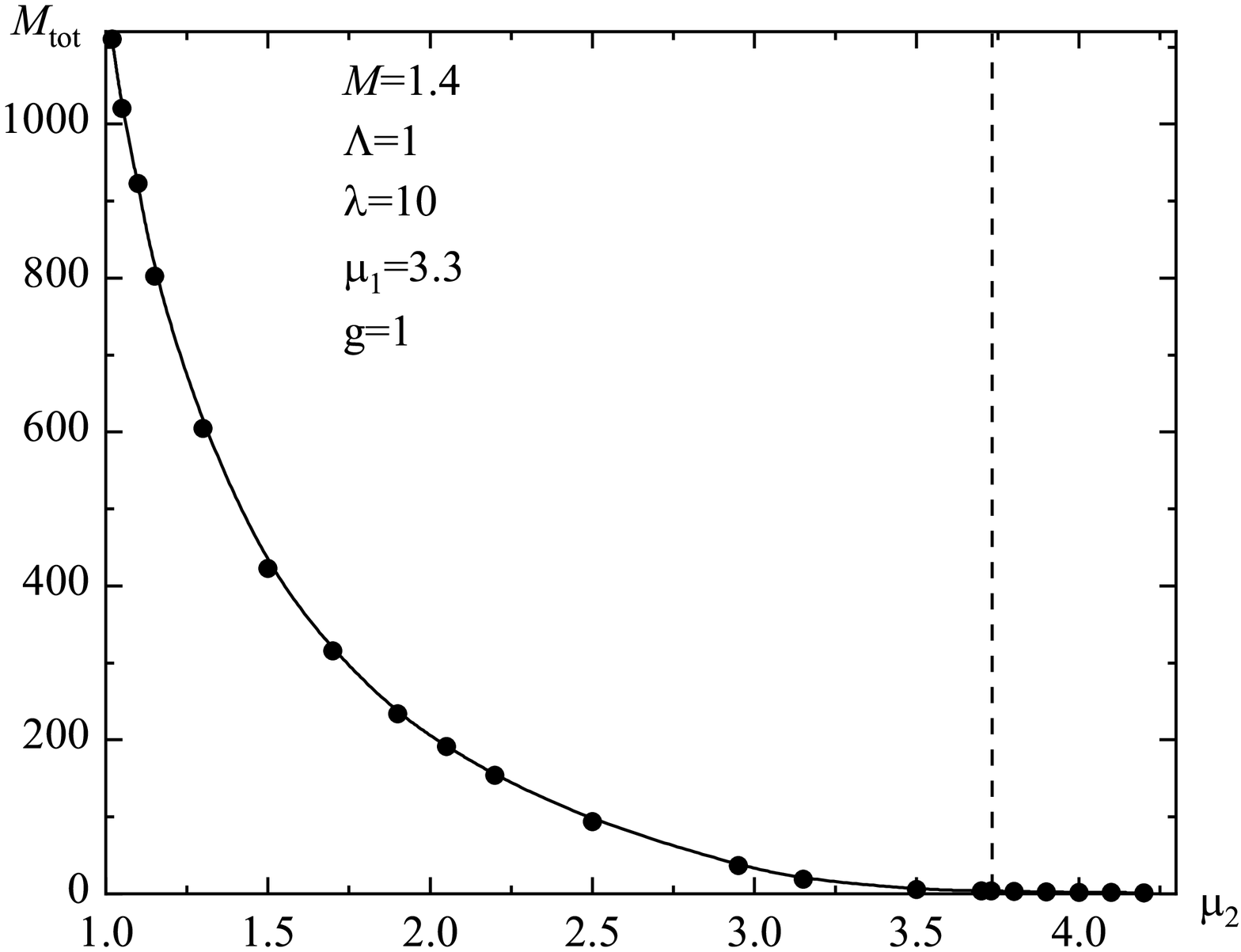}
%\vspace{-0.5cm}
\caption{The dependence of the total mass \eqref{M_tot} of the configurations under consideration on
$\mu_2$ for fixed values of the parameters  $M=1.4$, $\Lambda=1$, $\lambda=10$, $g=1$, and $\mu_1=3.3$.
The vertical dashed line separates the systems containing the electric and magnetic fields (located to the left of the line) from the configurations possessing only the magnetic field (located to the right of the line).
}
\label{fig_M_tot_mu_7_phi}
\end{figure}

In particular, Fig.~\ref{fig_fields_distr} shows the total energy density of the system under consideration obtained from~\eqref{0_40} using
\eqref{vec_pot}-\eqref{2_20} in the form
\begin{equation}
\begin{split}
	\varepsilon = & \frac{1}{g^2}
	\left[
	\frac{f_{, \theta}^2}{2 r^2}
	+ \frac{f_{, r}^2}{2}
	+ \frac{w_{, r}^2}{2}
	+ \frac{w_{, \theta}^2}{2 r^2}
	+ \cot \theta  \frac{w w_{, \theta}}{ r^2}
	+\frac{w w_{, r}}{r}
	+\frac{g^2}{2} \left(
		\frac{\phi_{, \theta}^2}{r^2} + \phi_{, r}^2
	\right) +
	\frac{f^2 w^2}{8}
	- \frac{\mu_1^2}{2}f^2 -
	\frac{ \mu_2^2 }{2}w^2
	\right.
\\
	&
	\left.
	+ \frac{\lambda}{2}
	\phi^2 \left(f^2 + w^2\right)
	 + \csc^2\theta \frac{w^2}{2 r^2} +
	\frac{ g^2 \Lambda}{4}
	\left(\phi^2 - M^2\right)^2
	\right].
\end{split}
\label{2_80}
\end{equation}
It is seen from the graphs of Fig.~\ref{fig_fields_distr} that the energy density is negative in some regions. To see
whether this will give a negative total energy of the system or not, we calculate the total mass (energy) of the configurations
under consideration,
\begin{equation}
M_{\text{tot}}=2\pi \int \varepsilon r^2 \sin{\theta}dr d\theta.
\label{M_tot}
\end{equation}
The results of calculations are given in Fig.~\ref{fig_M_tot_mu_7_phi} for the range $1.02\leq\mu_2\leq 4.2$
for fixed values of other system parameters. For $\mu_2=\mu_{2(\text{crit})}\approx3.73$, the function $f\to 0$, and with further increase of $\mu_2$,
it always remains equal to zero. Correspondingly, the configurations with $\mu_2 \lesssim \mu_{2(\text{crit})}$ possess both the magnetic and electric fields,
but for $\mu_2 \gtrsim \mu_{2(\text{crit})}$ only the magnetic field is present. Taking into account that the asymptotic expansions of the solutions at large  $r$
give the restriction on the mass $\mu_2 < \sqrt{\lambda M^2}$ (see below), the numerical calculations indicate that when
$\mu_2$ goes to this upper limit the functions $w\to 0$ (the magnetic field is switched off) and $\phi \to M$ over all space. Correspondingly,
the energy density and the total mass of the system vanish, as is seen from Fig.~\ref{fig_M_tot_mu_7_phi}.

\begin{figure}[h!]
\includegraphics[width=.7\linewidth]{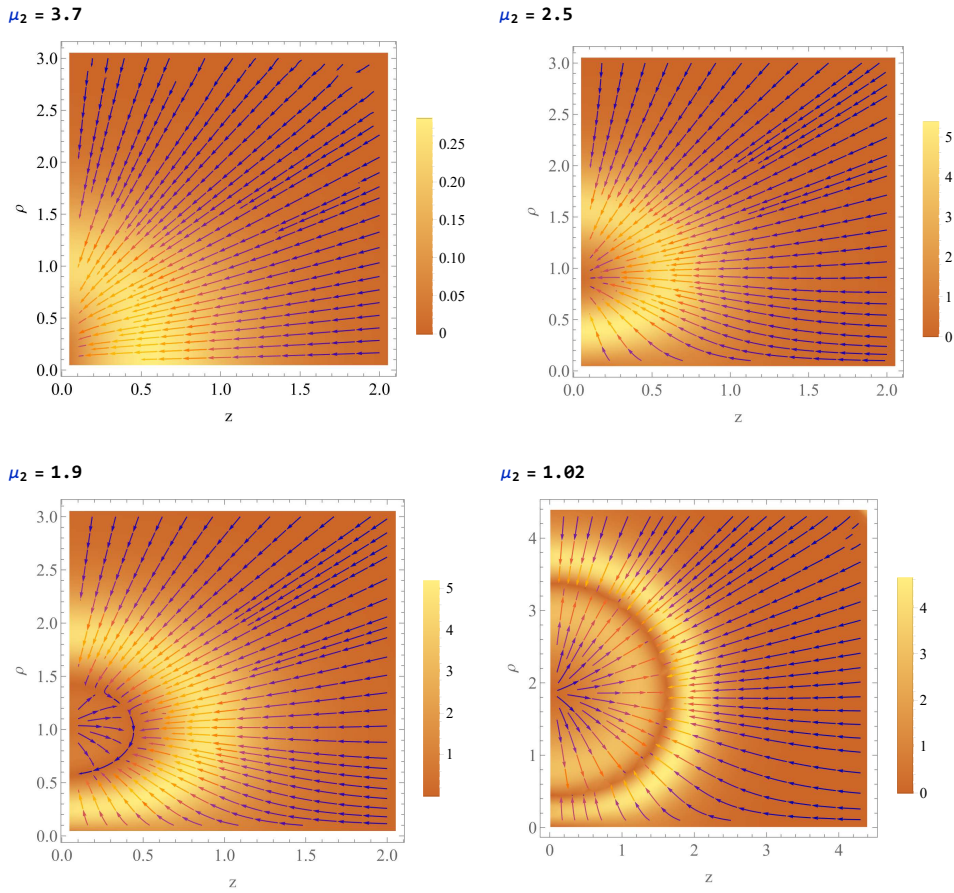}
%\vspace{-0.5cm}
\caption{The color electric field $\vec E^2$ strength distributions for different values of the  Proca mass $\mu_2$ and a fixed value
$\mu_1=3.3$. The parameters $M=1.4$, $\Lambda=1$, $\lambda=10$, $g=1$.
}
\label{fig_elect_stren_2}
\end{figure}

\begin{figure}[h!]
\includegraphics[width=.7\linewidth]{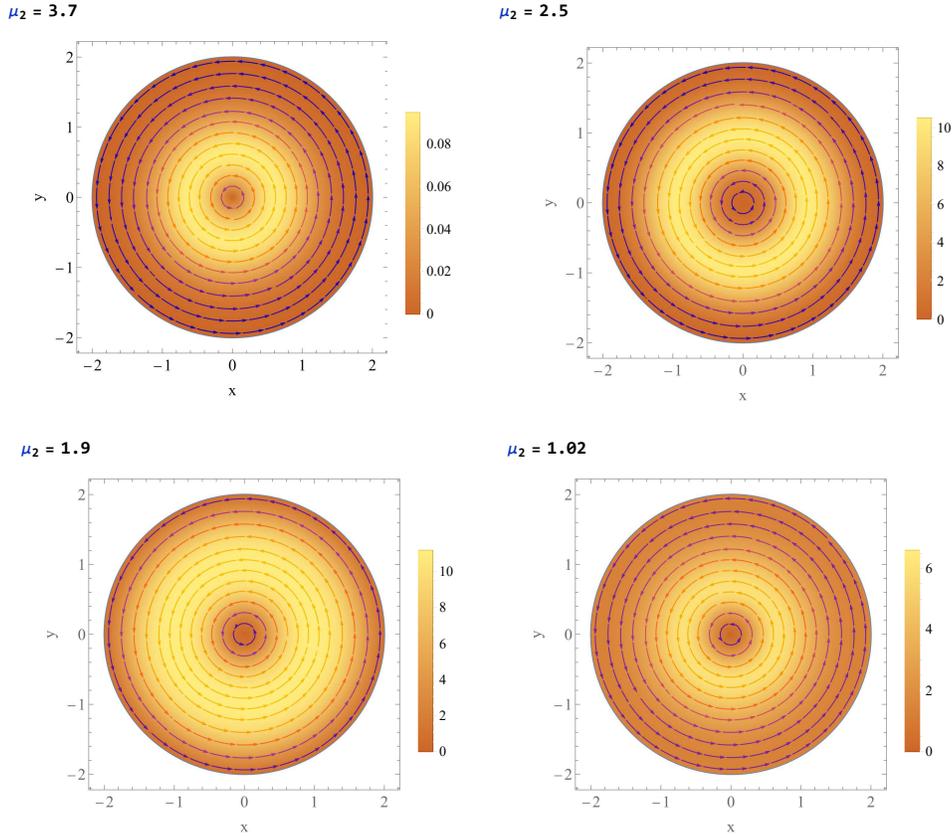}
%\vspace{-0.5cm}
\caption{The color electric field $\vec E^5$ strength distributions for different values of the  Proca mass $\mu_2$ and a fixed value
$\mu_1=3.3$. The parameters $M=1.4$, $\Lambda=1$, $\lambda=10$, $g=1$
(the graphs are plotted in the $\{x, y\}$ plane, i.e., when $\theta = \pi/2$).
}
\label{fig_elect_stren_5}
\end{figure}

\begin{figure}[h!]
\includegraphics[width=.7\linewidth]{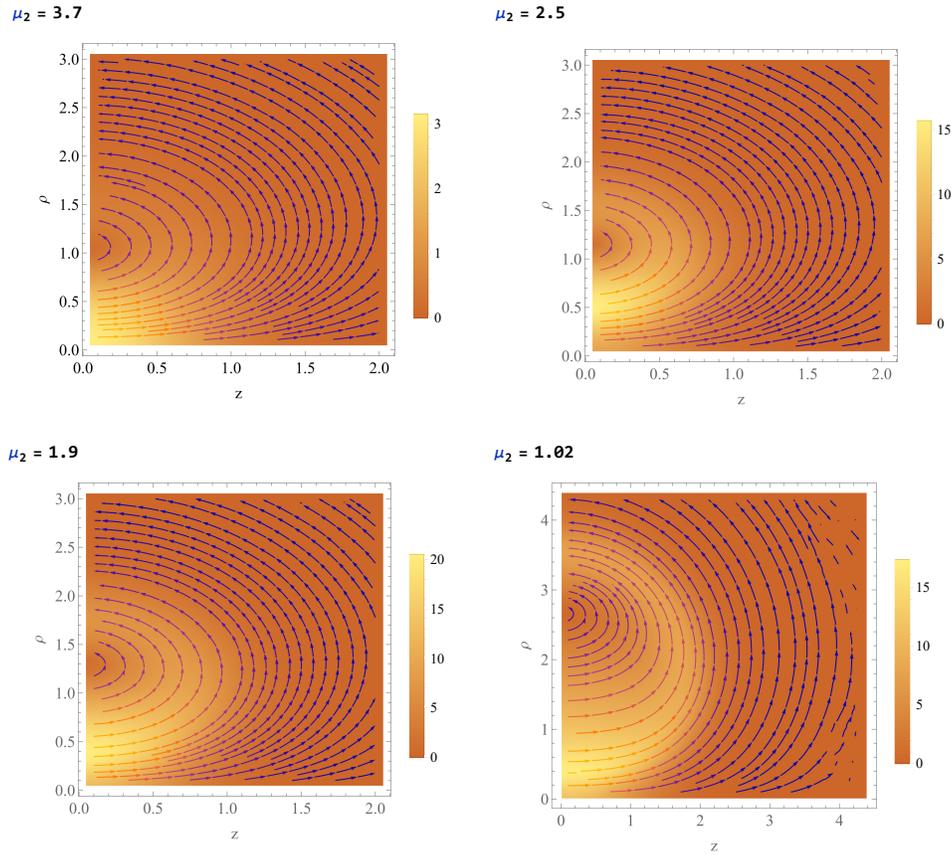}
%\vspace{-0.5cm}
\caption{The color magnetic field $\vec H^7$ strength distributions for different values of the  Proca mass $\mu_2$ and a fixed value
$\mu_1=3.3$. The parameters $M=1.4$, $\Lambda=1$, $\lambda=10$, $g=1$.
}
\label{fig_magn_stren}
\end{figure}

Using the solutions obtained, in Figs.~\ref{fig_elect_stren_2}-\ref{fig_magn_stren}, we have plotted the distributions of the color electric and magnetic fields for different values of $\mu_2$.
The analysis of these distributions indicates that
\begin{itemize}
\item According to Fig.~\ref{fig_elect_stren_2}, there is a point lying on a circle in the  $z=0$ plane
where the lines of force of the electric field $\vec E^2$ penetrate into. Aside from this, for sufficiently small values of
  $\mu_2$, there is a surface in the form of a torus where the lines of force of the electric field penetrate into as well.
\item As is seen from Fig.~\ref{fig_magn_stren}, the magnetic field $\vec H^7$
possesses a vortex centred at a circle in the $z = 0$ plane.
In this case, as $\rho \rightarrow 0$, the component $H^7_\rho \rightarrow 0$, whereas the component $H^7_z$ remains nonzero;
as a result, there is a flux of this field perpendicular to the transverse cross-section of the tube.
\item The configurations obtained for $\mu_2 \gtrsim \mu_{2(\text{crit})}$ resemble somewhat the Nielsen-Olesen tube:
both types of systems contain only a magnetic field with the flux directed along the $z$-axis.
\item For all the configurations obtained, the electric field $\vec E^5$ is a vortex field (see Fig.~\ref{fig_elect_stren_5}).
\end{itemize}


Thus, as in the case with an infinite tube of Sec.~\ref{inf_tube}, the finite tube under consideration does not contain
the flux of the longitudinal chromoelectric field $E^2_z$,
$$
	\Phi_z^E = 2 \pi \int \limits_0^\infty \rho E^2_z(z=0,\rho) d \rho =0,
$$
but possesses the flux of the longitudinal chromomagnetic field $H^7_z$,
$$
	\Phi_z^H = 2 \pi \int \limits_0^\infty \rho H^7_z(z=0,\rho) d \rho \neq 0.
$$
Also, the finite tube contains no energy flux density~-- the Poynting vector~\eqref{3_a_40} is zero.

Consider now the asymptotic behavior of the solutions found. As
$r \rightarrow \infty$, the fields $f, w \rightarrow 0$ exponentially fast; hence, one can neglect the nonlinear terms in Eqs.~\eqref{2_30}-\eqref{2_50}
containing $f^2$ and $w^2$. In turn, asymptotically, the field
$
	\phi \approx M - \eta \rightarrow M
$, and the function $\eta$ decays exponentially; this permits us to replace the term $\phi^2$ by $M^2$ in Eqs.~\eqref{2_30} and \eqref{2_40}.
As a result, from Eqs.~\eqref{2_30}-\eqref{2_50}, one can obtain the following asymptotic equations:
\begin{align}
	\bigtriangleup_{r, \theta} f +
	 \left( \mu_1^2 - \lambda M^2\right) f= & 0 ,
\label{2_90}\\
  \bigtriangleup_{r, \theta} w - \frac{w}{r^2\sin^2{\theta}}  +
	\left( \mu_2^2 -  \lambda M^2\right) w =& 0 ,
\label{2_100}\\
	  \bigtriangleup_{r, \theta} \eta -
	2  \Lambda M^2 \eta =& 0 ,
\label{2_110}
\end{align}
where $\bigtriangleup_{r, \theta}$ is the Laplacian operator. Eqs.~\eqref{2_90} and \eqref{2_110} have obvious solutions in the form
\begin{align}
	f \approx & C_{f}
	\left( Y\right)^0_{l_f}
	\frac{e^{- r \sqrt{\lambda M^2 - \mu_1^2}}}{r},
\label{2_120}\\
	\eta \approx & C_{\eta}
	\left( Y\right)^0_{l_\eta}
	\frac{e^{- r \sqrt{2 \lambda M^2}}}{r},
\label{2_130}
\end{align}
where $\left( Y\right)^0_{l_{f,\eta}}$ are spherical functions and $C_{f, \eta}$ are constants. In turn, Eq.~\eqref{2_100}
has a solution similar to \eqref{2_120}, but only with the angular part expressed in terms of special functions
(we do not show this expression here to avoid overburdening the text).
It follows from the above expressions that there are upper limits for the masses $\mu_1$ and $\mu_2$ ensuring
the exponential asymptotic decay of the solutions: $\mu_1^2, \mu_2^2< \lambda M^2$.

\subsection{The case with nonzero charges
}
\label{with_charge}

In this subsection we consider the case where the right-hand sides of the Proca equations \eqref{2_30} and \eqref{2_40}
contain nonzero current densities. When considering a self-consistent problem with a tube connecting quarks,
such currents must be created by spinor fields describing quarks. For the sake of simplicity, here we consider a  toy model
where the currents are given by hand. This means that the location of the quarks and the magnitude of color currents and charges created by them are fixed.

In the simplest case the currents can be given, for instance, by the Gaussian distribution,
\begin{equation}
	j^{a \mu} = \left( j_0\right)^{a \mu}
	e^{-\frac{R^2}{R^2_0}},
\label{1_300}
\end{equation}
where $\left( j_0\right)^{a \mu}$ is an arbitrary constant and $R$ is some function of $z$ and $\rho$.

\subsubsection{The case of nonzero charge density $j^{2 t}$ }

\begin{figure}[t]
\includegraphics[width=.7\linewidth]{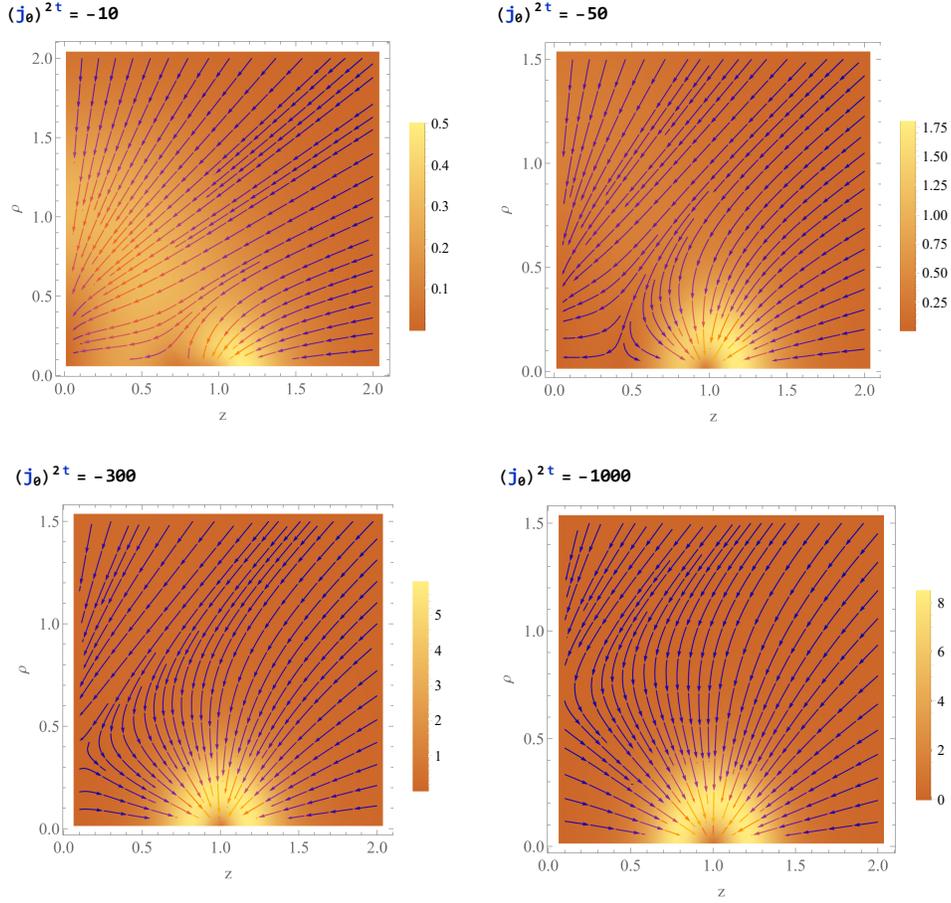}
%\vspace{-0.5cm}
\caption{The color electric field $\vec E^2$ strength distributions for different values of the  parameter $(j_0)^{2 t}$ and fixed values
 $\mu_1=3.3$ and $\mu_2=3.7$. The parameters $M=1.4$, $\Lambda=1$, $\lambda=10$, $g=1$, $R_0=0.2$, and $l=1$.
}
\label{fig_elect_stren_2_w_charge}
\end{figure}

Consider first the case where there is only the charge density $j^{2 t}$ in Eq.~\eqref{2_30}. For this case, we choose
 $	R^2 = \left( z - l \right)^2 + \rho^2$. Such a choice implies that  a ``quark'' is located on the tube axis at a distance
  $l$ from the origin of coordinates. Choosing different values of the constant
   $(j_0)^{2 t}$, in Fig.~\ref{fig_elect_stren_2_w_charge}, we have plotted the corresponding distributions of the color electric field $\vec E^2$.
As is seen from the figure, as $(j_0)^{2 t}$ increases (in modulus), the behavior of the lines of force changes qualitatively: if for small
$(j_0)^{2 t}$ the electric field created by the Proca field by itself is still comparable to the magnitude of the field determined by the charge,
it ceases to be so with increase of $(j_0)^{2 t}$ when the field determined by the charge becomes dominating;
this results in the aforementioned qualitative changes of the structure of the electric field.
This is illustrated by the appearance of a ``fixed'' point in the  $z = 0$ plane and at some distance from the tube axis.
The behavior of the lines of force is similar here to the behavior of phase trajectories for autonomous differential equations.

Distributions of the electric, $\vec E^5$, and magnetic, $\vec H^7$, fields change qualitatively only slightly  compared with the case without charge.
Hence, to avoid overburdening the text, we do not show them here.

\subsubsection{The case of nonzero current density $j^{7 \varphi}$ }

\begin{figure}[t]
\includegraphics[width=.7\linewidth]{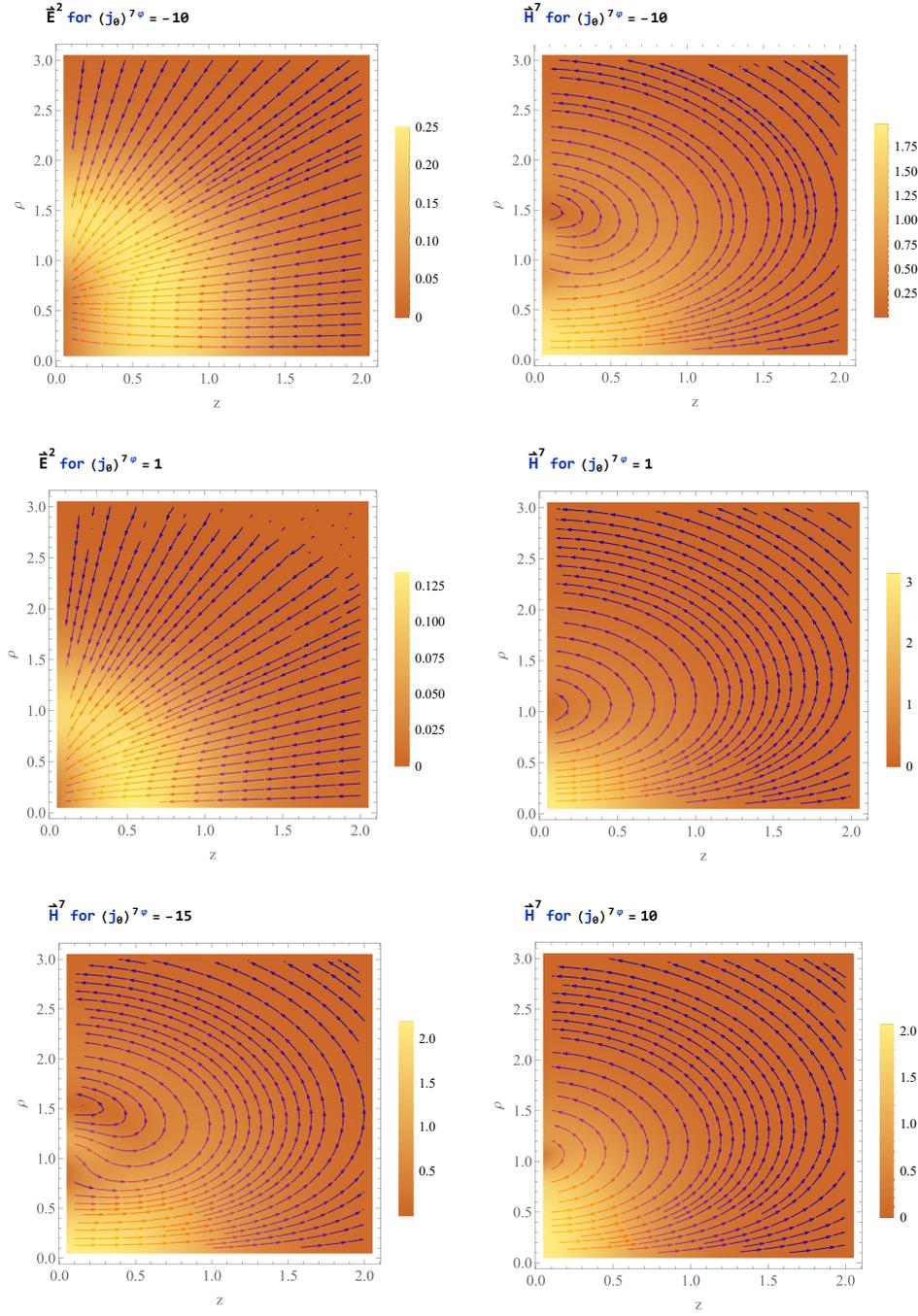}
%\vspace{-0.5cm}
\caption{The electric and magnetic fields strength distributions for different values of the  parameter $(j_0)^{7 \varphi}$
	and fixed values $\mu_1=3.3$ and $\mu_2=3.7$. The parameters $M=1.4$, $\Lambda=1$, $\lambda=10$, $g=1$, $R_0=0.2$, and $l=1$.
	For  $(j_0)^{7 \varphi}=-15$ and $10$, the electric field is absent.
}
\label{fig_elect_magn_with_current}
\end{figure}

Consider now the case where there is only the current density $j^{7 \varphi}$ in Eq.~\eqref{2_40}. For this case, we take
$R^2 = \left( \rho - l \right)^2 + z^2$ in Eq.~\eqref{1_300}; this implies the presence of the current of quarks  in the
 $z=0$ plane located on a circle of radius $l$. As in the case of the presence of the charge density considered above,
 we will change the value of the arbitrary constant $(j_0)^{7 \varphi}$ and keep track of the behavior of the electric and magnetic fields.
As a result, it turns out that both fields may exist simultaneously only in a very restricted range of values of
 $(j_0)^{7 \varphi}$. For example, for the same fixed values of the system parameters as those used in the case with a charge,
 there is the following range of possible values $-11\lesssim(j_0)^{7 \varphi}\lesssim 2$
 for which the system possesses both the electric and magnetic fields. When one goes beyond this range, only the magnetic field remains.
 Thus there are some critical values of $(j_0)^{7 \varphi}$ for which the electric field is switched off. 
 One might expect that for another values of the system parameters, distinct from those for which the aforementioned range of $(j_0)^{7 \varphi}$ has been found, 
 there will take place a similar situation where there will be some critical values of $(j_0)^{7 \varphi}$ determining the range in which configurations
 both with electric and magnetic fields do exist.
  
 The corresponding field strength distributions are given in Fig.~\ref{fig_elect_magn_with_current}.
 It is seen from them that the qualitative picture of the electric field strength distribution  changes only slightly (the same is true of
 the strength of the field $\vec E^5$ that is still a vortex field). In turn, the distribution of the magnetic field may depend considerably on
 the value of the constant $(j_0)^{7 \varphi}$; this especially manifests itself for large negative values of
$(j_0)^{7 \varphi}$ when the influence of the external source of the field becomes already comparable with the contribution caused by the
nonlinearity of the Proca fields.

\section{Discussion and conclusions}
\label{concl}

In the present paper, we have studied particlelike solutions with the flux of a magnetic field
in the non-Abelian Proca-Higgs theory. The crucial feature of these solutions is that they are localised
in space and have the flux of the color magnetic Proca field passing through the central cross-section of the Proca tube.
This allows the possibility of comparing the field configurations obtained with the Nielsen-Olesen tube.
In both cases, there are common features: (i)~the presence of a longitudinal magnetic field creating the flux of the magnetic field over the transverse cross-section of the tube; and
(ii)~the scalar field pushes out the magnetic field (the analogue of the Meissner effect). The main distinctions between the Proca tube and the Nielsen-Olesen tube are:
(i)~the Proca tube has finite size, whereas the Nielsen-Olesen tube is infinite; and (ii)~the Proca tube is a topologically trivial object.

Natural questions arising in this context are: (i)~Do Proca fields really exist in nature? (ii)~Where the tubes obtained by us could occur?
Regarding the first question, it should be pointed out that at the present time numerous studies are carried out to find such fields in nature,
see, e.g., Refs.~\cite{Bustillo:2020syj,Kase:2020yhw,Sanchis-Gual:2018oui,deFelice:2017paw}. According to another point of view,
Proca theories may be not fundamental but some phenomenological approaches to other theories. For instance, one might suppose that the
SU(2) Proca-Higgs theory could serve as a phenomenological description of some phenomena in SU(3) QCD.
For example, this could be in the case where some degrees of freedom belonging to the subgroup $\text{SU(2)}\subset \text{SU(3)}$ acquire mass as a result of quantization, and
such degrees of freedom are approximately described by a SU(2) Proca theory, whereas the remaining degrees of freedom are purely quantum and are approximately described
by a scalar field as a condensate. Of course, such a description is approximate and is rather similar to a phenomenological description of superconductivity by the Ginzburg-Landau equation.

Regarding the second question, one may suppose that if Proca fields do exist in nature (it is unimportant whether they are fundamental or phenomenological ones),
at high temperatures, in non-Abelian Proca theories, there can exist a ``quark-Proca-gluon-Higgs'' plasma; this is
analogous to what happens in QCD when the quark-gluon plasma occurs.

Summarizing the results obtained in the present paper:
\begin{itemize}
\item We have studied finite-size tube solutions in the non-Abelian Proca-Higgs theory both with and without external sources.
\item It is shown that such configurations have the finite total energy and the flux of the magnetic field over the transverse cross-section.
\item It is demonstrated that there exist some critical values of the system parameters $\mu_2$ and $(j_0)^{7 \varphi}$
for which the electric field is switched off and only the magnetic field remains nonzero.  
\item It is established that for the Proca tubes to exist, the presence of a scalar field is necessary;
such field acts as Cooper pairs in a superconductor and pushes out the Proca field creating localised objects.
\item Comparison of properties of the solutions obtained with the properties of the Nielsen-Olesen tube has been carried out.
\item It is assumed that such particlelike solutions may describe quasiparticles in the hypothetical ``quark-Proca-gluon-Higgs'' plasma.
\end{itemize}

In conclusion, we would like to emphasize the rich internal structure of Proca theories allowing the possibility of obtaining the particlelike solutions studied in the present paper,
whose existence is ensured by the presence of a mass of the vector Proca bosons.

\section*{Acknowledgements}

We gratefully acknowledge support provided by the program No.~BR10965191 of the Ministry of Education and Science of the Republic of kazakhstan.
We are also grateful to the Research Group Linkage Programme of the Alexander von Humboldt Foundation for the support of this research.
We thank Jutta Kunz and Burkhard Kleihaus for many useful discussions and helpful correspondence.


\begin{thebibliography}{99}

\bibitem{Brito:2015pxa}
  R.~Brito, V.~Cardoso, C.~A.~R.~Herdeiro, and E.~Radu,
``Proca stars: Gravitating Bose-Einstein condensates of massive spin 1 particles,''
  Phys.\ Lett.\ B {\bf 752}, 291 (2016).

\bibitem{Herdeiro:2017fhv}
  C.~A.~R.~Herdeiro, A.~M.~Pombo, and E.~Radu,
  ``Asymptotically flat scalar, Dirac and Proca stars: discrete vs. continuous families of solutions,''
  Phys.\ Lett.\ B {\bf 773}, 654 (2017).

\bibitem{Dzhunushaliev:2019kiy}
  V.~Dzhunushaliev and V.~Folomeev,
  ``Dirac star in the presence of Maxwell and Proca fields,''
  Phys.\ Rev.\ D {\bf 99}, no. 10, 104066 (2019).

\bibitem{Herdeiro:2019mbz}
   C.~Herdeiro, I.~Perapechka, E.~Radu, and Y.~Shnir,
  ``Asymptotically flat spinning scalar, Dirac and Proca stars,''
  Phys.\ Lett.\ B {\bf 797}, 134845 (2019).

%\cite{Bustillo:2020syj}
\bibitem{Bustillo:2020syj}
J.~C.~Bustillo, N.~Sanchis-Gual, A.~Torres-Forn\'e, J.~A.~Font, A.~Vajpeyi, R.~Smith, C.~Herdeiro, E.~Radu, and S.~H.~W.~Leong,
``GW190521 as a Merger of Proca Stars: A Potential New Vector Boson of $8.7\times 10^{-13}$  eV,''
Phys. Rev. Lett. \textbf{126}, no.8, 081101 (2021).

%\cite{Heisenberg:2017xda}
\bibitem{Heisenberg:2017xda}
L.~Heisenberg, R.~Kase, M.~Minamitsuji, and S.~Tsujikawa,
``Hairy black-hole solutions in generalized Proca theories,''
Phys. Rev. D \textbf{96}, no.8, 084049 (2017).

\bibitem{Allys:2015sht}
  E.~Allys, P.~Peter, and Y.~Rodriguez,
  ``Generalized Proca action for an Abelian vector field,''
  JCAP {\bf 1602} no.02,  004  (2016).

%\cite{Allys:2016kbq}
\bibitem{Allys:2016kbq}
E.~Allys, P.~Peter, and Y.~Rodriguez,
``Generalized SU(2) Proca Theory,''
Phys. Rev. D \textbf{94}, no.8, 084041 (2016).

%\cite{Minamitsuji:2016ydr}
\bibitem{Minamitsuji:2016ydr}
M.~Minamitsuji,
``Solutions in the generalized Proca theory with the nonminimal coupling to the Einstein tensor,''
Phys. Rev. D \textbf{94}, no.8, 084039 (2016).

%\cite{DeFelice:2016yws}
\bibitem{DeFelice:2016yws}
A.~De Felice, L.~Heisenberg, R.~Kase, S.~Mukohyama, S.~Tsujikawa, and Y.~l.~Zhang,
``Cosmology in generalized Proca theories,''
JCAP \textbf{06}, 048 (2016).

\bibitem{Dzhunushaliev:2019sxk}
V.~Dzhunushaliev and V.~Folomeev,
``Proca tubes with the flux of the longitudinal chromoelectric field and the energy flux/momentum density,''
Eur. Phys. J. C \textbf{80}, no.11, 1043 (2020).

\bibitem{Dzhunushaliev:2020eqa}
V.~Dzhunushaliev, V.~Folomeev, T.~Kozhamkulov, A.~Makhmudov, and T.~Ramazanov,
``Non-Abelian Proca theories with extra fields: particlelike and flux tube solutions,''
Phys. Scripta \textbf{95}, no.7, 074013 (2020).

\bibitem{Dzhunushaliev:2021uit}
V.~Dzhunushaliev, V.~Folomeev, and A.~Tlemisov,
``Linear energy density and the flux of an electric field in Proca tubes,''
Symmetry \textbf{13}, no.4, 640 (2021).

\bibitem{NielsenOlesen:1973}
H. B. Nielsen  and P. Olesen,
``Vortex-line models for dual strings,''
 Nucl. Phys. B \textbf{61}, 45 (1973).

%\cite{Shuryak:2003ty}
\bibitem{Shuryak:2003ty}
E.~V.~Shuryak and I.~Zahed,
``Rethinking the properties of the quark gluon plasma at T approximately T(c),''
Phys. Rev. C \textbf{70}, 021901 (2004).


%\cite{Shuryak:2004tx}
\bibitem{Shuryak:2004tx}
E.~V.~Shuryak and I.~Zahed,
``Towards a theory of binary bound states in the quark gluon plasma,''
Phys. Rev. D \textbf{70}, 054507 (2004).


\bibitem{fidisol}
W. Sch\"{o}nauer and R. Wei{\ss},
``Efficient vectorizable PDE solvers,''
J. Comput. Appl. Math. {\bf 27}, 279 (1989).

%\cite{Kase:2020yhw}
\bibitem{Kase:2020yhw}
R.~Kase, M.~Minamitsuji, and S.~Tsujikawa,
``Neutron stars with a generalized Proca hair and spontaneous vectorization,''
Phys. Rev. D \textbf{102}, no.2, 024067 (2020).

%\cite{Sanchis-Gual:2018oui}
\bibitem{Sanchis-Gual:2018oui}
N.~Sanchis-Gual, C.~Herdeiro, J.~A.~Font, E.~Radu, and F.~Di Giovanni,
``Head-on collisions and orbital mergers of Proca stars,''
Phys. Rev. D \textbf{99}, no.2, 024017 (2019).

%\cite{deFelice:2017paw}
\bibitem{deFelice:2017paw}
A.~de Felice, L.~Heisenberg, and S.~Tsujikawa,
``Observational constraints on generalized Proca theories,''
Phys. Rev. D \textbf{95}, no.12, 123540 (2017).

\end{thebibliography}
\end{document}